\documentclass[twocolumn,aps,showpacs,showkeys,prl,superscriptaddress,reprint]{revtex4-1}
\usepackage{graphicx}
\usepackage{amsmath}
\usepackage{amssymb}
\usepackage{amsfonts}
\usepackage{bm}
\usepackage{hyperref}
\usepackage{color}

\newcommand{\lz}{\langle \hat{L}_z \rangle}
\newcommand{\qmax}{q_\text{max}}

\begin{document}

\title{Boundaries for efficient use of electron vortex beams to measure magnetic properties}

\author{J\'{a}n Rusz}
\affiliation{Department of Physics and Astronomy, Uppsala University, P.O. Box 516, 75120 Uppsala, Sweden}
\affiliation{Institute of Physics, Czech Academy of Sciences, Na Slovance 2, 182 21 Prague, Czech Republic}
\author{Somnath Bhowmick}
\affiliation{Department of Physics and Astronomy, Uppsala University, P.O. Box 516, 75120 Uppsala, Sweden}
\affiliation{Department of Materials Science and Engineering, Indian Institute of Technology, Kanpur 208016, India}

\begin{abstract}
Development of experimental techniques for characterization of magnetic properties at high spatial resolution is essential for progress in miniaturization of magnetic devices, for example, in data storage media. Inelastic scattering of electron vortex beams (EVB) was recently reported to contain atom-specific magnetic information. We have developed a theoretical description of inelastic scattering of EVB on crystals and performed simulations for EVB of different diameters. We show that use of an EVB wider than an interatomic distance does not provide any advantage over an ordinary convergent beam without angular momentum. On the other hand, in the atomic resolution limit, electron energy loss spectra measured by EVB are strongly sensitive to the spin and orbital magnetic moments of studied matter, when channeling through or very close to the atomic columns. Our results demonstrate the boundaries for efficient use of EVB in measurement of magnetic properties.
\end{abstract}


\maketitle

For several decades, data storage technologies are in a tireless evolution to keep up with processing of ever-increasing amounts of data. The technology behind data storage relies to a large fraction on magnetic properties of materials. Reducing the dimensions of magnetic bits into nanometer scale naturally requires characterization techniques that provide the means to measure magnetic properties at desired spatial resolution. This resolution is slowly getting out of reach for x-ray based techniques, such as x-ray magnetic circular dichroism. In 2006 an analogous technique, but performed with transmission electron microscope, was discovered \cite{nature} -- the electron magnetic circular dichroism (EMCD). EMCD relates spin and orbital magnetic moments to a difference of electron energy loss spectra measured at specific crystal orientations. As an electron microscopy based technique it brought a promise of element-sensitive magnetic characterization at atomic resolution. Since then, EMCD went through a rapid development with significant improvements both in spatial resolution and signal to noise ratio \cite{lacbed, lacdif, emcd2nm, lsfollow}. Early adopters have successfully used it in their applications \cite{nanostuff, klie, bacteria, nanozno, nanofe3o4, cro2, fe3o4chan}. Yet, EMCD has not reached a stage of a wide spread as a routine characterization technique. The major obstacle is a low signal to noise ratio, which is due to the fact that EMCD needs to be measured on crystals at scattering directions between the transmitted beam and Bragg spots.

In an attempt to overcome these difficulties, Verbeeck et al.\ \cite{vortjo} have used electron vortex beams (EVB; \cite{uchida,mcmorran}) to measure an EMCD signal. This experiment suggests that EMCD can be measured at a transmitted beam, if the beam would carry an angular momentum. Provided we could obtain EVB with an intensity comparable to an ``ordinary'' convergent electron beam \cite{strongvortex}, this would lead to EMCD spectra with substantially enhanced signal to noise ratio. The recipe is simple: one should measure electron energy loss spectrum with an EVB with angular momentum $\lz{}=+\hbar$ and another one with an EVB with $\lz{}=-\hbar$, and their difference should provide an EMCD spectrum.

Theoretical developments have followed \cite{idrobo, vortcryst, lloydpra, lloydprl, vortat2, xin, lubk, yuan} and provided understanding of formation of EVBs, their elastic scattering on crystals and inelastic scattering on individual atoms. A common feature of these works is a focus on EMCD at an atomic resolution, which naturally demands atomic-size vortex beams. A question of using EVB for magnetic characterization at mesoscopic scale (about 1nm and beyond, in the present context) has not been studied, despite its great potential for applications. What is also missing is an understanding of an inelastic interaction of EVBs with matter---an assembly of atoms---a key question for applying EVB for EMCD measurements. From the experimental point of view, quite surprisingly, further works utilizing EVB for measurement of EMCD have not appeared in literature so far. Lack of follow-up experiments and an incomplete theoretical understanding motivated us to explore theoretically and computationally the inelastic scattering of EVB on magnetic materials.

In this Letter we develop a theory of inelastic scattering of EVB on matter. Using body-centered cubic crystal of iron as a benchmark structure, we show that sensitivity of the EVB to magnetic moments crucially depends on its diameter. We demonstrate that EVB is efficient for detection of EMCD only in the atomic-resolution limit, where it provides higher signal-to-noise ratio than \emph{intrinsic} EMCD method \cite{nature} relying on dynamical diffraction.

\begin{figure}[t]
\includegraphics[width=7.5cm]{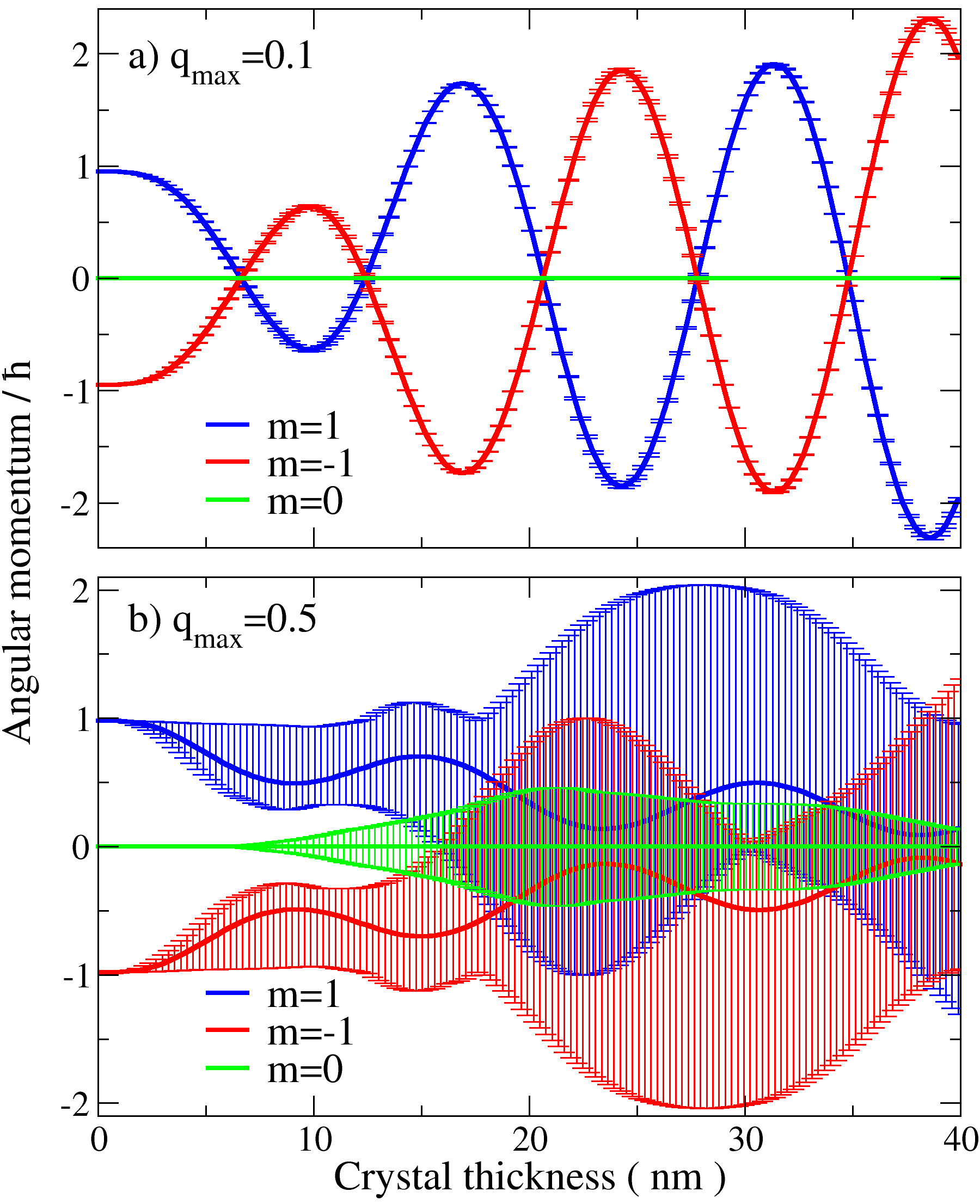}
\caption{Evolution of $\lz$ of the vortex beam as a function of sample thickness, averaged over different locations of the vortex core within a unit cell. ``Error bars'' indicate the spread of the angular momenta (minimum and maximum) at a given thickness. Top panel shows that a wide beam with $q_\text{max}=0.1$ have negligible spread and thus, virtually no dependence of angular momentum on the position of the vortex core. For a narrow vortex beam (bottom panel, $q_\text{max}=0.5$) there is a large spread of the angular momenta.}
  \label{fig:orbmom}
\end{figure}

The initial EVB wavefunction was generated in reciprocal space by $\phi(q,\varphi)=e^{im\varphi}\Theta(\qmax-q)$, where $q=\sqrt{k_x^2+k_y^2}$, $\varphi$ is the azimuthal angle, orbital angular momentum $\lz{}=\hbar m$, $\Theta$ is the Heaviside function and $\qmax$ determines the radius of the disk in the reciprocal space \cite{vortcryst}. We adopted two values of $\qmax$, namely 0.1~a.u.$^{-1}$, representing a beam much wider than one unit cell (referred as \emph{the wide beam}) and $\qmax{}=0.5$~a.u.$^{-1}$, representing a beam substantially narrower than the distance between the adjacent atomic columns in bcc-iron (\emph{the narrow beam}). It has been demonstrated that such atom-sized EVBs are within a reach \cite{vortatom,vortat2}. The initial wavefunction was propagated through a bcc-iron crystal along the $(001)$ direction up to a thickness of 40nm using a multislice method \cite{kirkland}, assuming an acceleration voltage of 200~keV. For both beam diameters, we have considered three values of angular momentum of the beam $\lz{}=-\hbar,0,\hbar$ and scanned the whole area of the unit cell by varying the lateral position of the beam center.

Development of the EVB angular momentum as a function of illumination spot and sample thickness was already studied in \cite{vortcryst}. We extend these results by considering two different beam diameters and showing the range of angular momenta that EVB can reach at a particular sample thickness, Fig.~\ref{fig:orbmom}. We find a dramatically different behavior of wide \emph{vs} narrow beam---forecasting our main result concerning the inelastic electron scattering further below.

The angular momentum of a wide beam is practically independent of the illumination spot (see minimum-maximum intervals in Fig.~\ref{fig:orbmom}a). This can be qualitatively predicted, knowing that the diameter of the beam covers several unit cells. In contrast, for a narrow beam the exchange of angular momentum between the beam and lattice is very sensitive to the illumination spot, as indicated by a large spread of values in Fig.~\ref{fig:orbmom}b.

In addition we note that, a beam with non-zero angular momentum can be obtained by propagating a narrow beam with $\lz{}=0$ through a crystal of suitable thickness, provided one can pass a narrow probe through an appropriate lateral position within the unit cell. Beyond 10nm, at certain illumination spots it acquires a non-negligible angular momentum, reaching a peak of value $\sim 0.5\hbar$ at a thickness of 20nm. However, an averaged value over the whole unit cell remains zero at all thicknesses (Fig.~\ref{fig:orbmom}b). For a beam with orbital angular momentum $\lz{}=\hbar$ the average over the unit cell does not vanish within the thickness range considered in our simulations. As in the case of a beam with $m=0$, it is possible to manipulate the probe's angular momentum by illuminating an appropriate spot in the unit cell and passing the beam through a sample of suitable thickness. Note that the range of accessible values is substantially enhanced compared to a probe with zero initial $\lz{}$.

The probe wavefunctions calculated by multislice method serve as an input for the inelastic electron scattering calculations \cite{bwconv}. We have employed the \emph{operator maps} technique \cite{opmaps} for evaluation of the inelastic scattering matrix elements of $L_3$ edge of bcc iron (energy loss 708~eV). This method allows to split the $L_3$-edge integrated inelastic scattering cross-section into a contribution due to holes in the $3d$-shell (referred as \emph{the non-magnetic signal}), and a contribution due to spin and orbital magnetic moment (i.e., EMCD integrated over $L_3$ edge; or \emph{the magnetic signal}). Technical details of the computational method will be reported elsewhere \cite{evbfollow}.

\begin{figure}[t]
  \includegraphics[width=7cm]{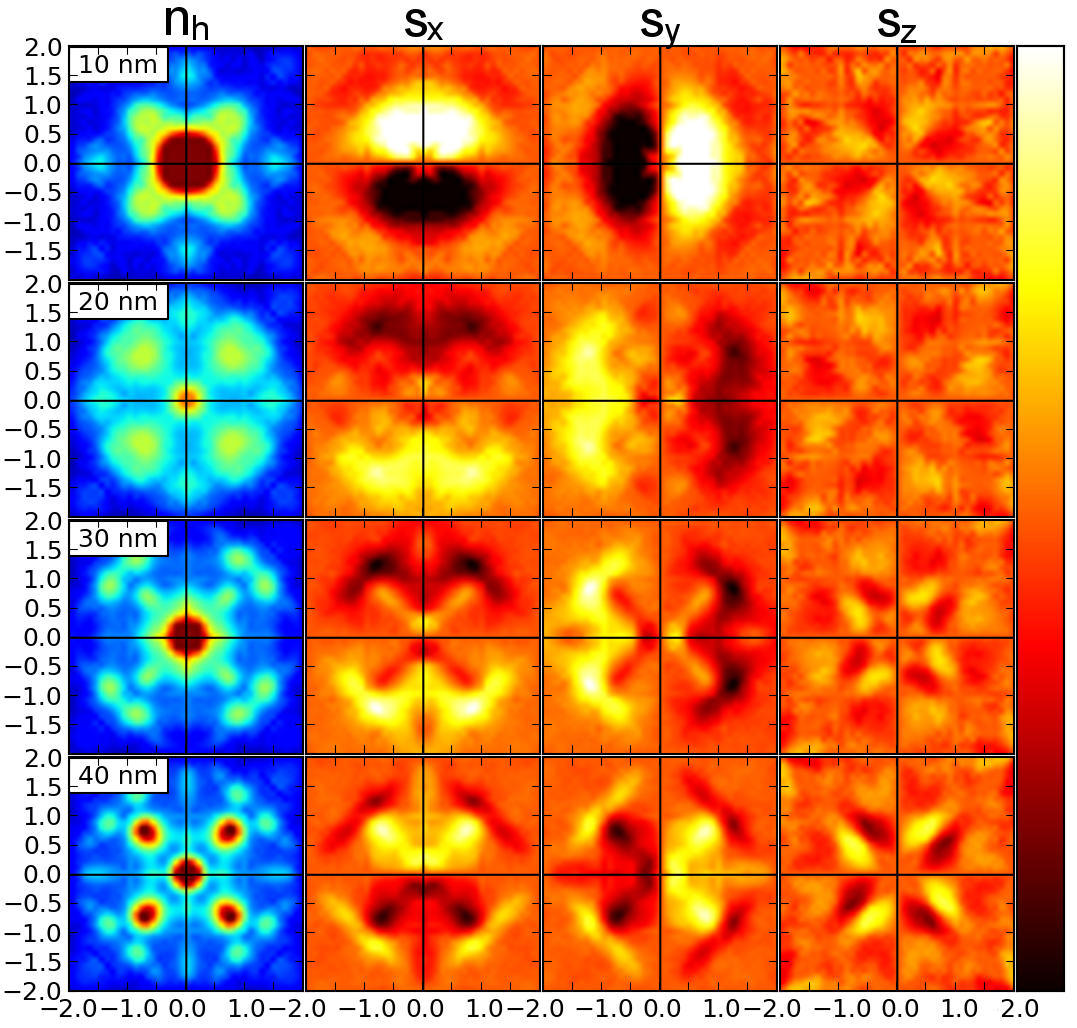}
  \caption{Inelastic scattering of a wide vortex beam. The grid of EFDIF patterns shows intensity per hole in $3d$ shell and per Bohr magneton of spin magnetization in $x,y,z$ direction (columns from left to right) for four different thicknesses, 10nm, 20nm, 30nm and 40nm (rows from top to bottom). Range of plots is from $-2G$ to $2G$, where $\mathbf{G}=(100)$. The color ranges are from 0 to 3.0 (blue to red) for the first column and -0.0625 to 0.0625 (black to yellow) for the second to fourth columns, respectively.}
  \label{fig:qmax01maps}
\end{figure}

\begin{figure}[t]
  \includegraphics[width=8.6cm]{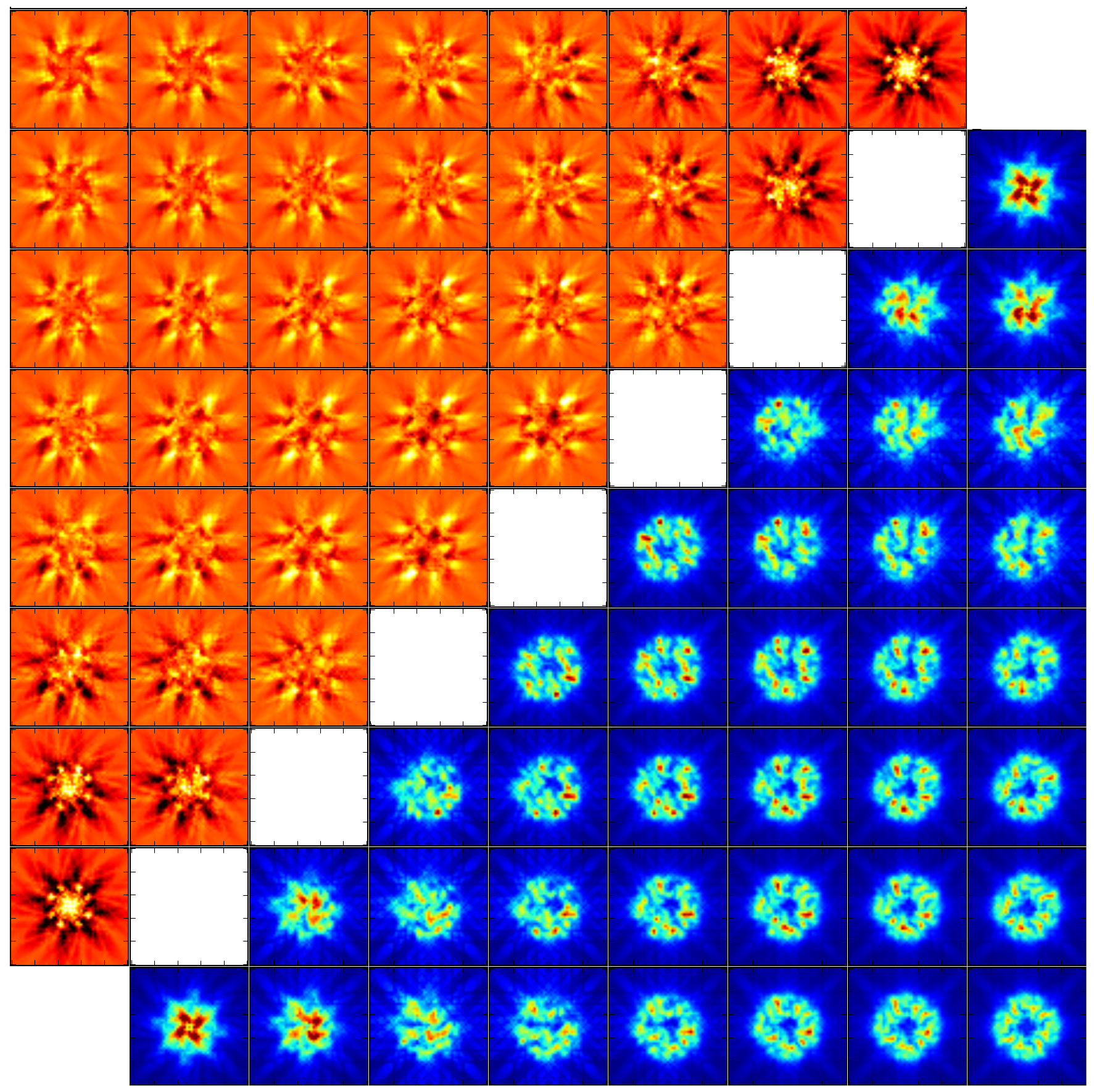}
  \includegraphics[width=8.5cm]{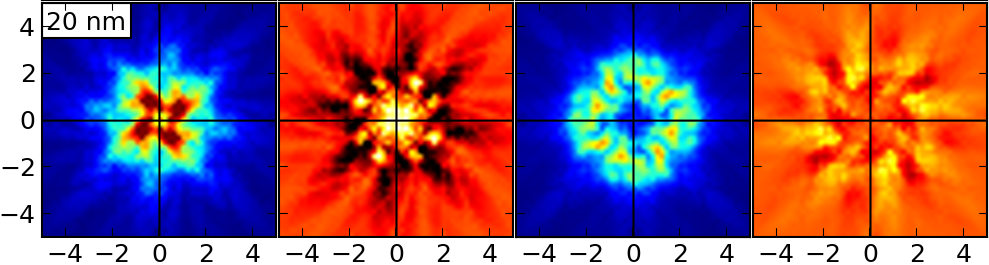}
  \caption{Inelastic scattering of a narrow vortex beam with angular momentum $\lz=+\hbar$, calculated for a sample thickness of 20nm. Figure displays a grid of calculated EFDIF patterns normalized per hole in $3d$ shell (right-bottom triangle) and per Bohr magneton of magnetization in $z$-direction (left-top triangle), representing the non-magnetic and magnetic signal, respectively. The color ranges are from 0 to 1.2 (dark blue to red) and -0.025 to 0.025 (black to yellow), respectively. The range of plots is from $-5G$ to $5G$ in both $x$ and $y$ directions, where $\mathbf{G}=(100)$. The maps in the lower left corner correspond to a vortex core passing through an atom at the origin of unit cell, while the maps in right top corner describe a vortex passing through a column of atoms in the centers of the body-centered cubic unit cell. Bottom panel shows the diffraction patterns (non-magnetic and magnetic part) for a vortex passing through an atomic column (left) and in between columns (right).}
  \label{fig:qmax05mapsp1}
\end{figure}

A striking result is obtained for the wide vortex beam, $\qmax{}=0.1$. Like in the case of exchange of angular momentum with lattice, the simulations show that the energy-filtered diffraction (EFDIF) patterns are independent of the position of the vortex center within the unit cell. Moreover, and this constitutes one of the main results of this Letter, these EFDIF patterns are independent of the angular momentum of EVB. In other words, for a wide vortex beam there is no influence of the beam vorticity on the observed diffraction patterns, which rules out the utility of EVB for measuring magnetic signal beyond atomic resolution. Representative EFDIF patterns are shown in Fig.~\ref{fig:qmax01maps}. Note that there is a non-negligible magnetic signal present in the diffraction plane for all three directions of magnetization. However, this signal originates solely from dynamical diffraction effects, i.e., it is an \emph{intrinsic} EMCD appearing due to the crystal itself acting as a beam-splitter \cite{nature}. In the light of these findings, we suggest that the EMCD signal observed by Verbeeck et al.\cite{vortjo} was of intrinsic origin.

For the narrow beam we observe rich and featureful dynamical diffraction effects. Inelastic scattering sensitively depends on the position of the vortex center within the unit cell, as is demonstrated in Fig.~\ref{fig:qmax05mapsp1}, showing EFDIF patterns for 36 positions of the EVB core from a triangular wedge mapping 1/8-th of the area of crystal unit cell. The development of the shape of the diffraction pattern is rather non-trivial, both for the non-magnetic and magnetic contribution. Magnetic signal is particularly strong when EVB passes close to the atom columns (bottom-left and top-right corner of Fig.~\ref{fig:qmax05mapsp1}).

\begin{figure}[t]
  \includegraphics[width=8.6cm]{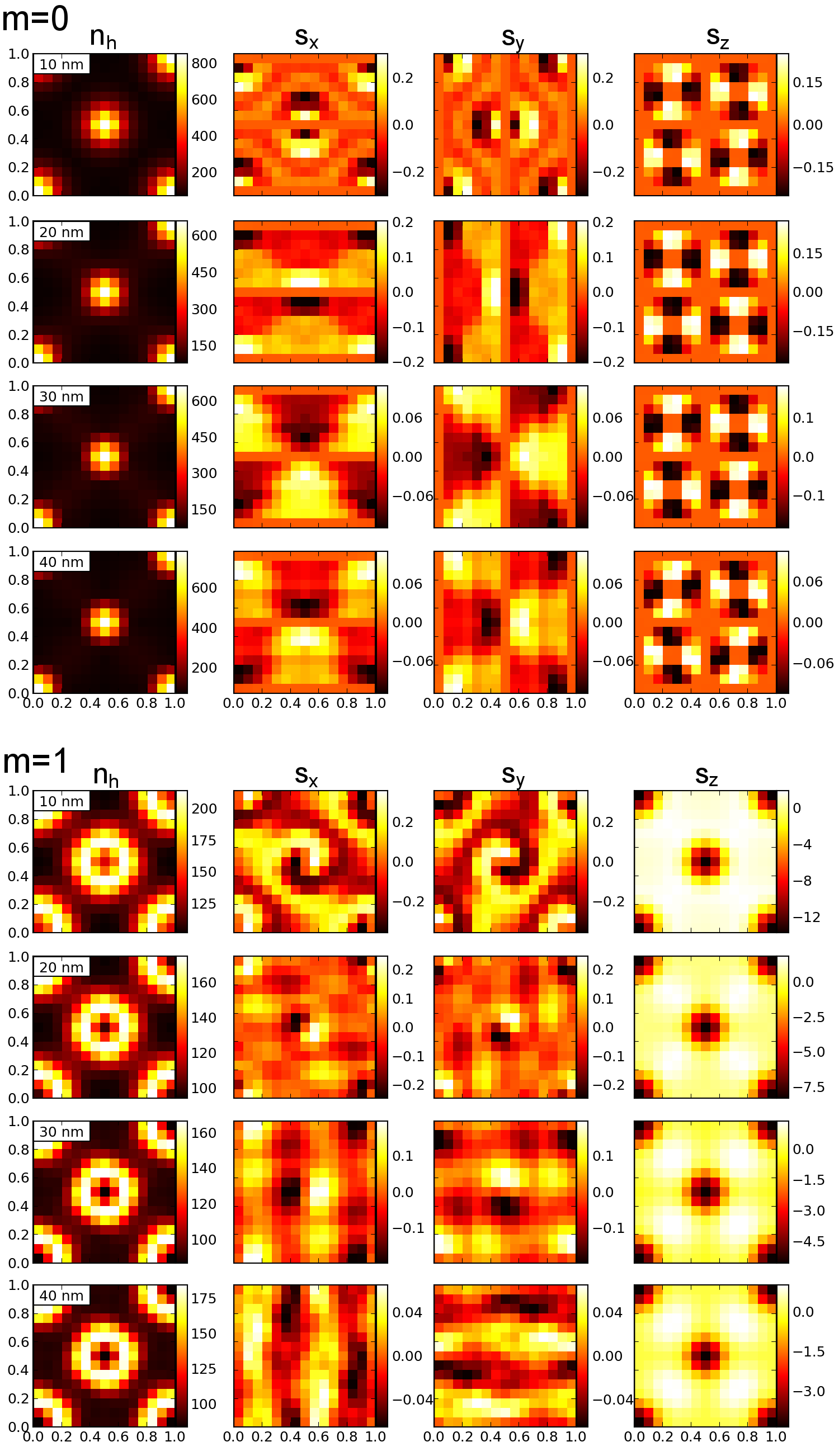}
  \caption{High-resolution energy-filtered images for a beam of zero angular momentum (top panel) and $\lz{}=\hbar$ (bottom panel). Individual rows correspond to thicknesses of 10nm, 20nm, 30nm and 40nm, respectively, and columns refer to a signal contributions normalized per hole or per $1\mu_B$ of spin magnetization along $x,y,z$ directions, respectively. Each pattern covers one unit cell.}
  \label{fig:hrefi}
\end{figure}

The EFDIF patterns as a function of illumination spot allow us to evaluate high-resolution energy-filtered images \cite{prange} (HR-EFI), which conveniently summarize the second main result of this Letter. We have simulated the detector aperture by a circle of radius $0.6G$ and $3.2G$ [$\mathbf{G}=(100)$] for the wide and narrow beam, respectively. These values are approximately equal to the $\qmax$ used to generate initial wavefunctions.

Calculations for the wide beam did not produce any contrast within the unit cell, as mentioned above. The non-magnetic signal is independent of a position of the EVB core and the magnetic signal vanishes after integration over an aperture---regardless of the angular momentum of the beam and magnetization direction.

For the narrow beam we have plotted the HR-EFI in Fig.~\ref{fig:hrefi}. Results for the beam with zero angular momentum show well-resolved positions of atomic columns. A non-zero magnetic signal can be detected, however it is of very low relative magnitude below 0.3\%. In the case of a vortex beam with $\lz{}=\hbar$ the maximum strength of the non-magnetic signal is reduced. It can be explained by a more spread beam of doughnut shape, which also leads to a lower spatial resolution---note the wider atomic columns in the non-magnetic component of HR-EFI for a vortex beam, compared to a beam with zero angular momentum (left columns of the upper and the lower panel of Fig.~\ref{fig:hrefi}). On the other hand, the magnetic signal originating from magnetic moment along $z$-direction is much more localized and significantly stronger than for a beam with $\lz{}=0$.

This shows that for a sufficiently narrow EVB channelling through an atomic column (in our case, within $0.6$\AA{} from the atomic column) the intensity of inelastically scattered EVB in the forward direction is substantially influenced by magnetic moments within the atomic column. At certain probe positions at 10nm the magnetic signal reaches 10\% of the maximum non-magnetic signal. In comparison to the intrinsic EMCD, where the net EMCD signal in the case of bcc-Fe does not exceed 0.5\% of the transmitted beam intensity \cite{mcremcd}, this is a significantly stronger signal and remains stronger within the studied range of thicknesses up to 40nm.

Importantly, an integral of HR-EFI over the area of whole unit cell provides very weak magnetic signals of the order of less than 0.05\% of the non-magnetic signal. Therefore our calculations demonstrate that in order to utilize EVB for measurement of EMCD one has to scan the crystal at an atomic resolution. Only in this case there is a theoretical possibility to measure enhanced magnetic signal with EVB, but this is yet to be demonstrated experimentally.

In conclusion, we have demonstrated the range of applicability of EVB for measuring magnetic properties of matter. Our results should stimulate further development of EVB experiments at an atomic resolution, which could become the method of choice for element-specific magnetic characterization of thin crystalline layers.

J.R. acknowledges Swedish Research Council, G\"{o}ran Gustafsson's Foundation, Swedish National Infrastructure for Computing (NSC center) and computer cluster \textsc{Dorje} at Czech Academy of Sciences.

\end{document}